# `osl-ephys`: A Python toolbox for the analysis of electrophysiology data

Mats W.J. van Es[1], Chetan Gohil[1,2], Andrew J. Quinn[1,3], Mark W. Woolrich[1]

[1] Oxford Centre for Human Brain Activity, Wellcome Centre for Integrative Neuroimaging, Department of Psychiatry, University of Oxford, Oxford, United Kingdom.
[2] Brain and Mind Centre, School of Medical Sciences, University of Sydney, Sydney, Australia
[3] Centre for Human Brain Health, School of Psychology, University of Birmingham, Birmingham, United Kingdom

## Abstract

We describe OHBA Software Library for the analysis of electrophysiological data (`osl-ephys`). This toolbox builds on top of the widely used MNE-Python package and provides unique analysis tools for magneto-/electro-encephalography (M/EEG) sensor and source space analysis, which can be used modularly. In particular, it facilitates processing large amounts of data using batch parallel processing, with high standards for reproducibility through a config API and log keeping, and efficient quality assurance by producing HTML processing reports. It also provides new functionality for doing coregistration, source reconstruction and parcellation in volumetric space, allowing for an alternative pipeline that avoids the need for surface-based processing, e.g., through the use of Fieldtrip. Here, we introduce `osl-ephys` by presenting examples applied to a publicly available M/EEG data (the multimodal faces dataset). `osl-ephys` is open-source software distributed on the Apache License and available as a Python package through PyPi and GitHub.

## 1. Introduction

The analysis of neuroimaging data typically involves a series of complicated analysis steps which are deployed heterogeneously to suit both the dataset and the scientific question. In non-invasive human electrophysiology data, particularly magnetoencephalography (MEG) and electroencephalography (EEG), these steps include but are not limited to: preprocessing to clean the raw recordings, co-registration with other data modalities (e.g., Polhemus, MRI), source reconstruction, and a plethora of (mass) univariate or multivariate statistical analyses.

To this aim, the field has traditionally relied on a suite of open-source software toolboxes developed by individual research groups or community efforts (Delorme and Makeig, 2004; Litvak et al., 2011; Oostenveld et al., 2011; Tadel et al., 2011; Jenkinson et al., 2012; Gramfort, 2013; Lopez-Calderon and Luck, 2014; OHBA Analysis Group, 2014). However, most of these rely on licensed, third-party software like MATLAB (The MathWorks Inc., 2020), which makes it costly and limits processing of large amounts of data. Because of this, there is a current shift in the field to adopt the Python programming language and multiple Python packages specifically designed for the analysis of electrophysiology data have recently been published (Gramfort, 2013; Schirrmeister et al., 2017; Lu, 2020; Sabbagh et al., 2020; Brodbeck et al., 2023; Gohil et al., 2023; Jas et al., 2023; Ågren, 2023). Of these, MNE-Python (Gramfort, 2013; Larson et al., 2023) is by far the most widely adopted.

The field also sees an increasing use of publicly available and large datasets. At the same time, there are higher requirements for transparency and reproducibility by academic journals. Therefore, analysis tools must adapt to help meet these needs. One solution is MNE-Python's MNE-BIDS-pipeline, which offers automated processing of M/EEG data in BIDS format through a configutarion API. However, specification of the configuration is arguably non-trivial, and the pipeline does not allow for full analysis flexibility.

Here, we present the `osl-ephys` toolbox, a free and open-source Python package for the analysis of electrophysiology data, and part of the *OHBA Software Library (OSL)*. It is built on, and augments, the MNE-Python toolbox (Gramfort, 2013; Larson et al., 2023), and is developed with the following core principles:
1. Efficient processing of large amounts of data;
2. A concise *configuration* API that summarizes a reproducible processing pipeline, and is easy to specify, interpret, and share;
3. Automatic generation of log files and HTML processing reports to enable reproducibility and provide quality assurance.
4. A modular setup, to facilitate integration with MNE-Python and other, third-party toolboxes in Python, or other programming languages (e.g., MATLAB).

Moreover, `osl-ephys` contains unique processing functions for analysing M/EEG data, including FSL-based (Freesurfer independent) coregistration and volumetric source reconstruction pipeline (Jenkinson et al., 2012).

`osl-ephys` has a user friendly Application Programming Interface (API) based on the specification of a `config` object, which concisely holds all the information needed for the processing steps to apply. This helps ensure reproducibility and facilitates the processing of large amounts of data. Below we briefly outline the architecture of `osl-ephys` and the API, before presenting examples of each module.

## 2. Method

### 2.1 Overview

`osl-ephys` uses MNE-Python as a backbone. For example, data classes like `Raw`, `Epochs`, `Evoked`, `Info`, and other classes are directly adopted and used, and M/EEG data derivatives are typically saved as `fif` files. This allows for a seamless integration of `osl-ephys` with MNE-Python, or with other Python and MATLAB based toolboxes. It is straightforward for a user to a subset of the data processing pipeline in MNE-Python or other analysis toolboxes, while switching to and from `osl-ephys`.

Because the field of cognitive neuroscience is inherently multidisciplinary, individual researchers often lack formal training in programming. Therefore, usability and reproducibility are at the heart of `osl-ephys`'s design philosophy. In particular, in the two main `osl-ephys` modules (`preprocessing` and `source_recon`, see below) a processing pipeline is defined in the form of a comprehensive configuration (*"`config`"*) API. This is a Python dictionary that specifies the call to individual functions and the parameter settings for each. The user will typically specify the `config` as a string or `YAML` file, which `osl-ephys` will convert into a dictionary. During the processing, `osl-ephys` handles data bookkeeping and other complexities behind the scenes. A feature of the `config` is that it is easily shareable and can be easily used to reproduce analyses.

In the two main `osl-ephys` modules, the user typically interacts with high level pipeline "*chain*" functions (e.g., `run_proc_chain` and `run_src_chain` for the `preprocessing` and `source_recon` modules, respectively), or the "*batch*" equivalent (`run_proc_batch` and `run_src_batch`, respectively) if they want to process multiple datasets at a time by efficiently looping the chain function over the datasets (see *2.3 Batch Processing*). These functions take as main input arguments the `config`, and the input and output directories.

The `osl-ephys` *chain* functions load the data (i.e., using MNE-Python's data loaders) and call the appropriate function from the `config` dictionary together with the parameter arguments. In addition, these functions can create log files and HTML reports (see below). The log files improve reproducibility by writing information on processing steps, random seed, etc. The reports summarise the processing and can guide quality assurance.

```python
import osl_ephys

config = """
  meta:
    event_codes:
      fixation_cross: 1
  preproc:
    - notch_filter: {freqs: 50 100}
    - resample: {sfreq: 250}
    - bad_segments: {segment_len: 500}

"""

osl_ephys.preprocessing.run_proc_chain(
    config,
    input = "raw.fif",
    subject = "subject001",
    outdir = "processed/",
)
```

```python
import osl_ephys

def manual_func(dataset, userargs):
    ...
    return dataset

config = """
  meta:
    event_codes:
      fixation_cross: 1
  preproc:
    - notch_filter: {freqs: 50 100}
    - resample: {sfreq: 250}
    - bad_segments: {segment_len: 500}
    - manual_func: {}
"""

osl_ephys.preprocessing.run_proc_chain(
    config,
    input = "raw.fif",
    subject = "subject001",
    outdir = "processed/",
    extra_funcs=[manual_func]
)
```

Listing 1a. Structure of `osl-ephys` processing in the `preprocessing` module. The `config` specifies the processing recipe applied to the `input`. Preprocessed data are saved in the `processed/subject001` directory. Note that the source reconstruction module works similarly (See *3. Examples*).

Listing 1b. As in 1b, but now including a custom written function. The function is defined at the top, and takes as input `dataset` (containing the MNE-Python objects, e.g., `Raw`), and a `userargs` dictionary (containing the function variable specifications). The function is included in the config, and supplied to `run_proc_chain` as a list of extra functions in the `extra_funcs` input variable.

## 2.2 Modules

The `osl-ephys` toolbox contains several *modules* that provide tools for a specific analysis goal, e.g., preprocessing, source reconstruction, general linear modelling (GLM), etc. These modules are designed such that they can be flexibly combined with third party toolboxes.

### 2.2.1 Maxfilter

The Maxfilter module contains wrappers to the licensed MaxFilter™ (MEGIN, Finland) software, and this module can thus only be used on computers that have this software installed and licensed. Beside the standard call to MaxFilter™ with the optional input arguments, the module also contains functions that call MaxFilter™ multiple times in series, as is commonly used in the field (e.g., a bad channel detection step, a Signal Space Separation (Taulu et al., 2005) step, and transformation to a different coordinate frame, also see (Henson, 2024)).

### 2.2.2 Preprocessing

The `preprocessing` module contains functionality to pre-process M/EEG data. A simple example is given in Listing 1a. Preprocessing can include (wrapper) functions for e.g., selecting data, filtering, resampling, bad channel/segment detection, Independent Component Analysis, and applying MaxFilter™). It can also include more advanced analysis steps, e.g., spectral analysis using multi-tapers. Beside the plethora of functions offered by MNE-Python, this module also contains `osl-ephys` specific functions, e.g., identifying bad channels/segments using a generalized ESD test (Rosner, 1983), and removing zeroed out data caused by MaxFilter™. User defined functions can also be used directly in `osl-ephys` to offer fully flexbile pipelines (see Listing 1b and *2.2.4 Custom functions*).

The main user functions in the preprocessing module are the pipeline functions `run_proc_chain`, and `run_proc_batch`. These load in the input data and represents them in a `dataset` dictionary as `dataset[`raw`]`. Other derivatives generated during preprocessing will be added to the `dataset` dictionary, which is ultimately returned by the *chain/batch* functions and individual items are saved to disk.

In addition to the above, the `preprocessing` module contains command line functionality for interactive labelling of bad ICs using a graphical user interface (GUI). See *3. Examples*.

### 2.2.3 Source Reconstruction

The `source_recon` module contains tools for coregistration, volumetric source reconstruction, and for working with source space data more generally. This module differs contains more unique functionalities not present in MNE-Python. In particular, `osl-ephys` provides a source reconstruction pipeline that allows for coregistration and volumetric source reconstruction. This makes use of FSL to compute the anatomical surfaces (Jenkinson et al., 2012), and does not require any use of Freesurfer.

Note that the `source_recon` module still uses low-level MNE-Python functions on the backend wherever possible. For example, instead of returning specific `SourceEstimate` classes as in MNE-Python, `osl-ephys` returns MNE-Python `Raw` classes with source estimated data, such that further processing of the data can be done analogously to sensor space data.

Typically, the steps used in the FSL-based coregistration and volumetric source reconstruction pipeline are as follows:

1. Compute surfaces. Extract the inner skull, outer skin (scalp) and brain surfaces from structural, T1 weighted MRI (sMRI) data, using FSL (Jenkinson et al., 2012).
2. Coregistration. Coregisters the M/EEG data, head digitization points (i.e., Polhemus), and sMRI data. This is done using *Registration using Headshapes Including Nose in `osl-ephys`* (RHINO), analogous to the MATLAB implementation (OHBA Analysis Group, 2014). It consists of calculating a linear, affine transform from native sMRI space to Polhemus (head) space, using headshape points that optionally include the nose. RHINO firsts registers the Polhemus-derived fiducials (nasion, right and left pre-auricular points) in Polhemus space to the sMRI-derived fiducials in native sMRI space. RHINO then refines this by making use of Polhemus-derived headshape points that trace out the surface of the head (scalp), and ideally include the nose, using the iterative closest point (ICP) algorithm. Finally, these Polhemus-derived headshape points in Polhemus space are registered to the sMRI-derived scalp surface in native sMRI space.
3. Forward modelling. This is a wrapper for `mne.make_forward_solution`.
4. Source modelling. Typically done using a volumetric LCMV beamformer (Van Veen et al., 1997), but other methods will be implemented in the future (e.g. Minimum Norm Estimate).

Source reconstruction to the volumetric dipole grid can then be followed by use of a pre-defined parcellation to extract parcel time courses, and including the reduction of spatial leakage and correction for sign ambiguities:

5. Parcellation. Parcellates a dipole grid of source estimates by taking the principal component (or spatial basis set) of all dipoles in a parcel from a chosen (Nifti) parcellation. Multiple standard parcellation templates are supplied (e.g., AAL (Tzourio-Mazoyer et al., 2002)), as well as fMRI-derived parcellations optimized for M/EEG data (i.e., with a lower amount of parcels to match the rank of M/EEG data; e.g., Giles39 (Colclough et al., 2015), and Glasser52 (Kohl et al., 2023)).

6. Symmetric orthogonalization. Ambiguities in the source reconstruction of M/EEG data can cause spurious correlations between source time courses. We use symmetric multivariate leakage reduction (Colclough et al., 2015) to correct for these artificial correlations between a set of multiple regions of interest (i.e., parcels). Note that by being multivariate, this also corrects for so-called "ghost interactions" or "inherited connections" (Colclough et al., 2015; Palva et al., 2018).
7. Sign flipping. Ambiguities in the source reconstruction and calculation of parcel time courses result in an ambiguity in the sign, or polarity, of the parcel time courses. We adjust the sign/polarity of the parcel time courses using the assumption that we expect correspondence in the functional connectomes from different subjects. As such, we maximize the correlation of the covariance matrices between subjects.

As with the `preprocessing` module, the `source_recon` module has a high-level pipeline function (`run_src_chain`, and for batches of data: `run_src_batch`) that work with a `config` API, and can optionally generate log files and HTML reports.

### 2.2.4 Custom functions

It is in general possible to supply the *chain* and *batch* functions with custom written functions, defined by the user. These can be readily supplied to `osl-ephys` by specifying the function in the `config` and supplying the list of custom functions to the `extra_funcs` input variable in the *chain/batch* function. The only requirement is that they adhere to the following structure (this is slightly different for `preprocessing` and `source_recon` modules):

`preprocessing` *module*
- The function must take `dataset` and `userargs` as inputs
- Any options for the functions that are specified in the config can be retrieved from `userargs` (i.e., using `userargs.pop()`)
- The function must return `dataset`
- Any key in `dataset` can be manipulated, either in place or by adding a new key. New keys are saved by default in the subject folder, by replacing "`preproc_raw`" in the preprocessed data filename by the name of the new key.

`source_recon` *module*
- The function can take any user-defined input variables as input. The inputs to `run_src_chain/batch`: `outdir, subject, preproc_file, smri_file, epoch_file, reportdir, logsdir,` are always passed to the custom function.
- Changes must be directly saved to disk, rather than returning function outputs.

### 2.2.5 General Linear Model

The `glm` module contains data classes and functions that combine functions from MNE-Python, custom code, and Python packages for linear modelling (Quinn, 2019; Quinn and Hymers, 2020) into modality specific (e.g. M/EEG data) tools for linear modelling. This includes the ability to do confound modelling and hierarchical modelling, including for spectral analysis (i.e., instead of the commonly used Welch periodogram, where the average is used; (Quinn et al., 2024)), and significance testing via non-parametric statistics. Additionally, it contains functions for visualizing (statistically significant) effects.

The GLM functionality cannot be directly added to a config for the `preprocessing` or `source_recon` pipeline function, because of the added complexity of specifying a design matrix. These tools are typically applied in a separate Python script or by specifying a custom function that is supplied to the *preprocessing* `config` (see *3. Examples*).

### 2.2.6 Utilities

The `utils` module contains several helpful utility functions that can be directly deployed by the user, and/or is used in some of the higher level `osl-ephys` functions. Current utility functions include data loaders (for e.g., OPM-MEG, SPM data), and functions for parallel processing (see *2.3 Batch processing*), logging, simulation, and file handling, see *3. Examples*.

### 2.2.7 Report

The `report` module allows for the generation of interactive HTML reports to gain insights into the analysis carried out by the high-level pipeline functions and guide quality assurance (QA). Separate reports are generated by default when using the `preproc` and `source_recon` *batch* processing functions, but can also be created manually. They include a subject report for in depth information about individual subjects/session, and a summary report summarising measures all across subjects/session.

The subject report contains information about each recording (e.g., number of sensors recorded, duration of the recording), and the analysis in the form of data tables and (interactive) figures, e.g., showing which channels were marked as "bad", the result of the coregistration, etc. These figures are generated in a *reports* directory, in subdirectories for each individual subject/session, together with a `data.pkl` file. This file contains symbolic links to the appropriate figures, as well as plain text and numeric information about the session (e.g., also including a copy of the text from the log files).

The summary report gives an overview of the processing pipeline, including custom function definitions, and has interactive data tables summarizing various metrics from all subject/session reports. This can guide the user to have a detailed look at specific subject report, for example for those subjects that had excessive number of channels marked as "bad", or high errors in the coregistration.

## 2.3 Batch processing

Essential in the design philosophy is the ability to process large batches of data efficiently. `osl-ephys`'s high-level pipeline functions therefore have `batch` function counterparts (e.g., `run_proc_batch` and `run_src_batch`), which take in a `config` and lists of file paths (or a path to a text file containing the file paths) for the data to be processed. We integrated `Dask` (Dask Development Team, 2016) for parallel processing these batches of data efficiently, using as many computational resources (i.e., CPU cores) as the user has available.

### 2.4 Log files

The *chain* and *batch* functions create log files to keep track of all the functions and configuration options that were applied to the data, and the output they generated. The log files also include the random seed, which improves reproducibility of the pipeline (a global random seed can also be set manually). Separate log files are created for each subject/session, and a separate batch log is also created when using the *batch* functions. The subject/session logs are also appended to `dataset[`raw`].extras[`description`]`, so a processed file will always contain a history of the functions applied to it.

### 2.5 Examples

We use the publicly available multimodal faces dataset (Wakeman and Henson, 2015), v0.1.1 available on OpenfMRI (Poldrack and Gorgolewski, 2017), to illustrate the use of `osl-ephys`. This dataset contains data of 19 subjects, each of which participated in six MEG recording sessions. The subjects engaged in a visual perception task where they saw a series of famous, novel, familiar (repetitions of novel faces) faces, and scrambled faces. To ensure participants were paying attention, participants had to indicate whether faces were symmetrical or asymmetrical with a button press. We analyse these data in a typical analysis workflow that optimally demonstrates the use cases and API of `osl-ephys`; the research question is not based on scientific novelty. Concretely, we first preprocess the MEG data in sensor space, and then reconstruct the sources and combine them into 52 parcels (Kohl et al., 2023). Next, we epoch the parcel time courses, and use a first level GLM to contrast real faces minus scrambled faces, separately for each session. We then model the group effect of this contrast in a second level GLM. All scripts used for this analysis are available on GitHub (https://github.com/OHBA-analysis/osl-ephys/tree/main/examples/toolbox-paper). HTML reports and log files are available on OSF (https://osf.io/2rnyg/).

### 2.6 Documentation

Documentation is available on readthedocs (https://osl-ephys.readthedocs.io/en/latest/). This includes installation instructions, a full list of function references (API), and tutorials. Source code is available on GitHub (https://github.com/OHBA-analysis/osl-ephys/tree/main).

### 2.7 Development

`osl-ephys` is under active development, and community contributions are welcome on the GitHub page (https://github.com/OHBA-analysis/osl-ephys/tree/main), in the form of GitHub Issues and pull requests. New `osl-ephys` versions will be released on GitHub and PiPy when significant changes in the toolbox have beent made.

### 2.8 Citing `osl-ephys`

For the most up to date information on how to cite `osl-ephys` read the `CITATION` file on GitHub (https://github.com/OHBA-analysis/osl-ephys/blob/main/CITATION.cff).

# 3. Results

Below, we show the results of using `osl-ephys` on the multimodal faces dataset with a typical analysis pipeline, including:
    1) preprocessing,
    2) coregistration and source reconstruction
    3) epoching and first-level general linear modelling (GLM)
    4) second level (group) statistical analysis using a GLM.

The results shown here are based on `osl-ephys 1.1.0`, `mne 1.3.1`, `fslpy 3.11.3`, and `Python 3.8.16`.

## 3.1 Batch preprocessing

Listing 2 shows how the `osl-ephys` preprocessing pipeline is setup in the `__main__` body of a Python script (Listing 2), which is necessary in order to use `Dask` for parallel processing. This first specifies the inputs to `osl-ephys.preprocessing.run_proc_batch` and the preprocessing `config`. The `config` contains a *meta* section specifying the event codes and names for each event, and a *preproc* section with each preprocessing step that will be run in turn. The preprocessing steps are by default applied to `dataset[`raw`]` (i.e., the input data; unless they are specific to a different MNE-Python class, e.g., `Epochs`). The appropriate function is found by matching the function name to 1) any custom written functions (supplied to the `extra_funcs` parameter), 2) `osl-ephys` specific functions, and 3) MNE-Python methods on `Raw`, and `Epochs` classes.

The first 5 steps of the *preproc* section are MNE-Python functions directly applied to the `Raw` data object. `bad_segments` and `bad_channels` are `osl-ephys` unique functions that use of a generalized ESD (gESD) test (Rosner, 1983) to do bad segment/channel detection, where the `mode: diff` option applies the gESD test to the temporal derivative of the channel time courses. `ica_raw` and `ica_autoreject` are MNE-Python-wrappers with `osl-ephys` specific default options (e.g., applying a 1 Hz high-pass filter before ICA).

Next, the paths to the raw MEG data are specified using the `Study` class. This enables data paths to be specified using multiple wild cars, and selects existing paths that satisfy the wild cards (here, all paths are selected, but, for example, one can select only the sessions of e.g., subject 1 using `study.get(sub_id=1)`). Session identifiers are then specified that will be used as subdirectory to save each session's data in, as well as the base of the saved data (see below), and a generic output directory. Lastly, parallel computation is enabled using `Dask` (having already specified the *Dask* `Client`). Running Listing 2 creates the output structure and derivatives in Figure 1.

```python
import os
from dask.distributed import Client

from osl_ephys import preprocessing, utils

if __name__ == "__main__":
    client = Client(n_workers=16, threads_per_worker=1) # specify to enable parallel processing
    basedir = "ds117"

    config = """
      meta:
        event_codes:
          famous/first: 5
          famous/immediate: 6
          famous/last: 7
          unfamiliar/first: 13
          unfamiliar/immediate: 14
          unfamiliar/last: 15
          scrambled/first: 17
          scrambled/immediate: 18
          scrambled/last: 19
      preproc:
        - find_events: {min_duration: 0.005}
        - set_channel_types: {EEG061: eog, EEG062: eog, EEG063: ecg}
        - filter: {l_freq: 0.5, h_freq: 125, method: iir, iir_params: {order: 5, ftype: butter}}
        - notch_filter: {freqs: 50 100}
        - resample: {sfreq: 250}
        - bad_segments: {segment_len: 500, picks: mag}
        - bad_segments: {segment_len: 500, picks: grad}
        - bad_segments: {segment_len: 500, picks: mag, mode: diff}
        - bad_segments: {segment_len: 500, picks: grad, mode: diff}
        - bad_channels: {picks: mag, significance_level: 0.1}
        - bad_channels: {picks: grad, significance_level: 0.1}
        - ica_raw: {picks: meg, n_components: 40}
        - ica_autoreject: {picks: meg, ecgmethod: correlation, eogmethod: correlation,
                           eogthreshold: 0.35, apply: False}
        - interpolate_bads: {reset_bads: False}
    """

    # Study utils enables selection of existing paths using various wild cards
    study = utils.Study(os.path.join(basedir, "sub{sub_id}/MEG/run_{run_id}_raw.fif"))
    inputs = sorted(study.get())

    # specify session names and output directory
    subjects = [f"sub{i+1:03d}-run{j+1:02d}" for i in range(19) for j in range(6)]
    outdir = os.path.join(basedir, "processed")

    preprocessing.run_proc_batch(
        config,
        inputs,
        subjects,
        outdir,
        dask_client=True,
    )
```

Listing 2. `osl-ephys` batch preprocessing script.

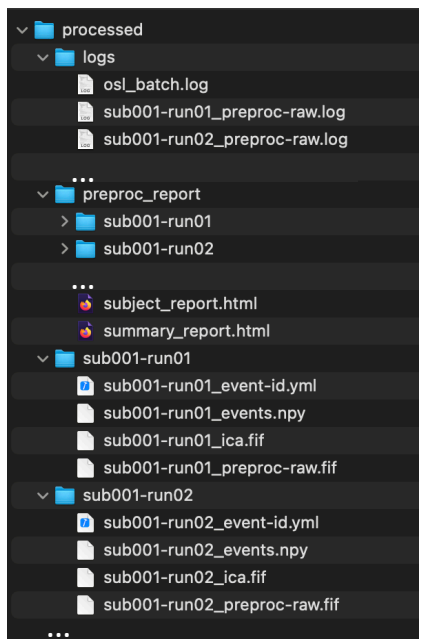

Figure 1. The output directory structure of `run_proc_batch`. All outputs are saved in the general output directory specified in the function call ("processed"). Within this, a sub directory is created for each subject/session that contains the preprocessed data, as well as *logs*, and *preproc_report* directories, containing the relevant files for all subjects.

### 3.2 Logs and preprocessing report

The `osl_proc_batch` function has additionally generated log files for each session and a batch log file. The batch log documents high level information about the batch processing including time stamps, the random seed, the `config`, and in how many files preprocessing was successful. Each session specific log file documents all processing steps applied to the data, including other relevant function outputs (e.g., the number of bad channels detected). The log files thus provide detailed information and can aid in reproducibility of the pipeline.

`osl-ephys` generates reports that can be used for quality assurance (QA), and which can easily be shared. It contains two HTML files: `subject_report.html` (Figure 2) and `summary_report.html` (Figure 3). The subject report presents detailed qualitative and quantitative metrics of the preprocessing applied to each subject (e.g., general data info, number of bad channels/segments detected, power spectra, etc.), whereas the summary report summarizes the batch preprocessing. It includes a table with summary metrics for each subject (e.g., percentage of data marked as "bad", number of bad channels, number of ECG-related ICs, etc.). This table is interactive and can guide the user to individual subjects/sessions which need to be manually checked. This is especially useful when a large amount of data is processed and manually checking each subject is not feasible.

For example, sorting the table based on *Bad ICA (total)* reveals there were 27 bad ICs detected in *sub008-run03*, of which 26 were labelled EOG. The subject report shows that most of these are spurious, and thus this dataset requires extra attention, for example by changing the preprocessing options or manually adapting the labels (see *3.3 Manual ICA labelling*). The reports also contain the batch and subject logs respectively, and if present, the summary report contains error logs for files that returned errors during preprocessing.

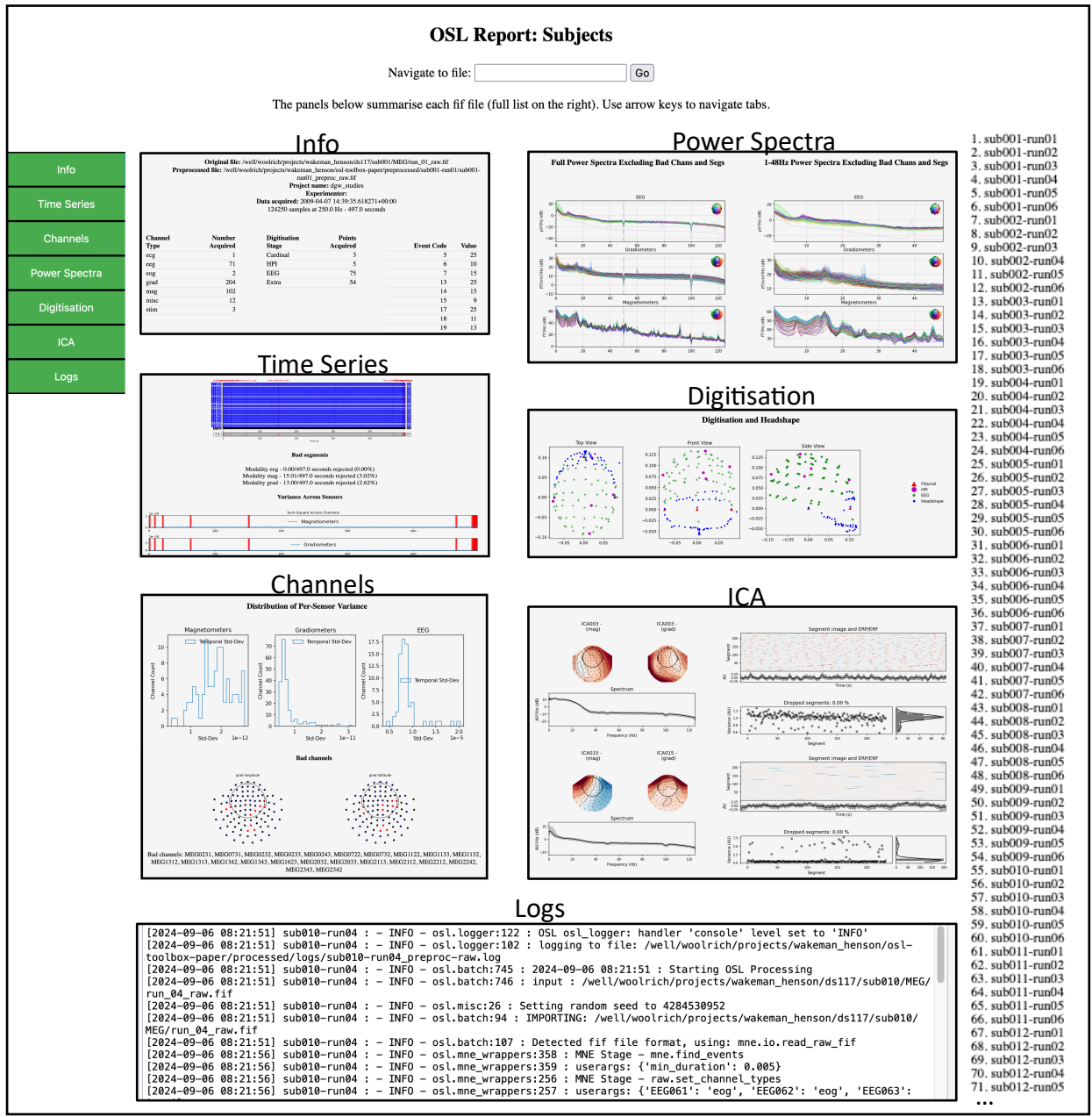

Figure 2. Example of the `preprocessing` subject report. This HTML page contains tabs for different aspects of quality assurance (QA) for each subject/session processed by `run_proc_batch`. The user can browse between tabs on the left for each subject/session in the list on the right. Each tab contains quantitative and qualitative information regarding the preprocessing output.

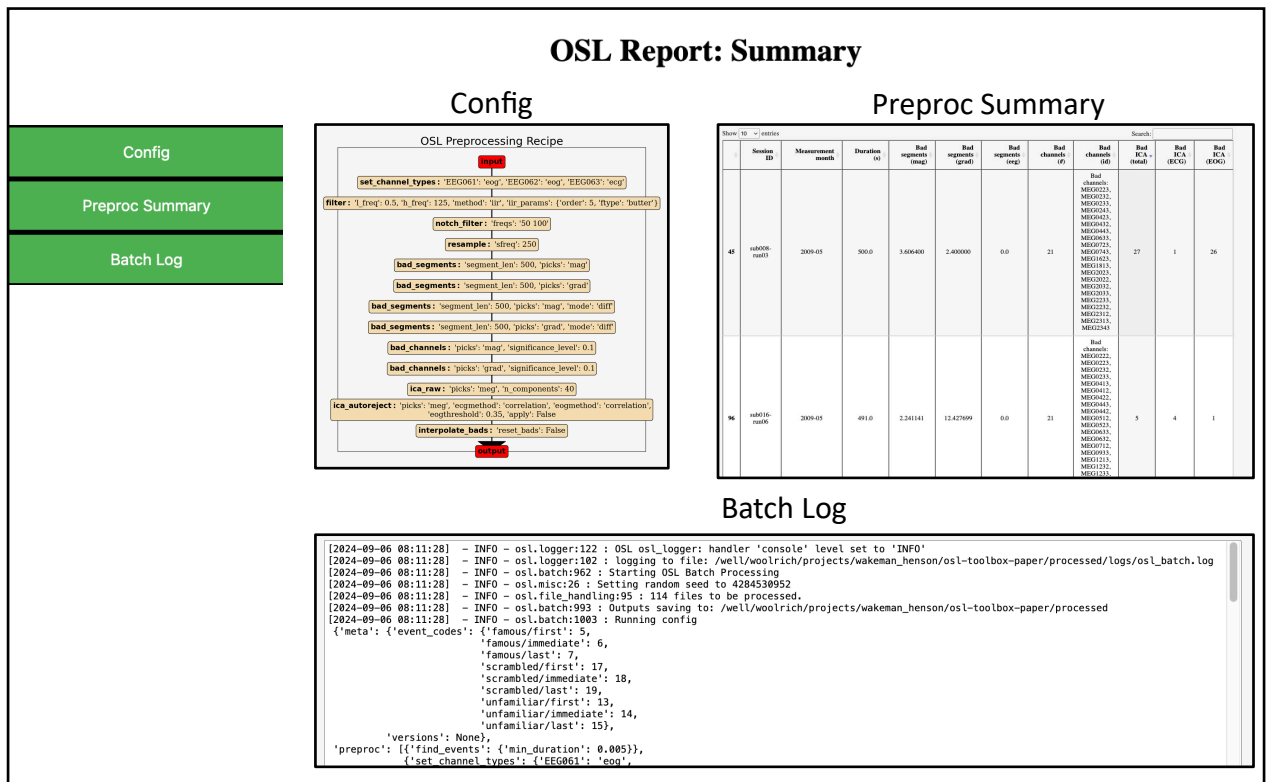

Figure 3. Example of the `preprocessing` summary report. Contains summary information and quantitative metrics of all files processed using `run_proc_batch`. The *Preproc Summary* table is interactive and can guide the user to specific subjects/sessions which might require further attention.

### 3.3 Manual ICA rejection

In MEG preprocessing, ICA is typically used to remove cardiac- and ocular-related artefacts. The ICs that capture these physiological artefacts can potentially be identified using the correlation with the electrocardiogram (ECG) and electrooculogram (EOG) time series, if these were recorded. If they are not, or if they are not of sufficient quality, manual inspection might be necessary. A combination of automatic and manual detection is recommended, i.e., manually refining the automatic first-pass labelling based on the `osl-ephys` report.

During batch preprocessing, ICA is run and identifies ICs with high correlations with the EOG/ECG signals. However, these ICs have not yet removed from the MEG data. As mentioned above, many ICs were spuriously labelled in *sub008-run03*. The user can use the interactive labelling tool `osl_ica_label` to manually correct the selection of bad ICs. It can be called from the command line in an active `osl-ephys` environment (i.e., `osle`) with only a handful required inputs, explained below:

```
(osle) > osl_ica_label reject_option preproc_dir session_name
```

- `reject_option`: indicating which of the ICs should be removed from the data. Can be `all` (i.e., automatically and manually labelled ICs), `manual` (i.e., only manually labelled ICs), or `None` (i.e., save the `ICA` object but do not remove any components from the data).

- `preproc_dir`: general output directory, i.e., the same as supplied to `run_proc_batch`.
- `session_name`: subject/session specific identifier, i.e., the same as supplied to `run_proc_batch`.

In this example, no components have yet been removed from the data. First, the selected components need to be manually checked for a few sessions. Therefore, the command line call is as follows:

```
(osle) > osl_ica_label None processed sub008-ses03
```

This opens the interactive tool (Figure 4), which shows the IC weights and time courses (and ECG/EOG time courses at the bottom. The user can browse through ICs (vertical scroll bar) and time (horizontal scroll bar) and (de-)select ICs where appropriate, using button pressed to optionally label selected ICs as correlate of artefact types indicated on the right.

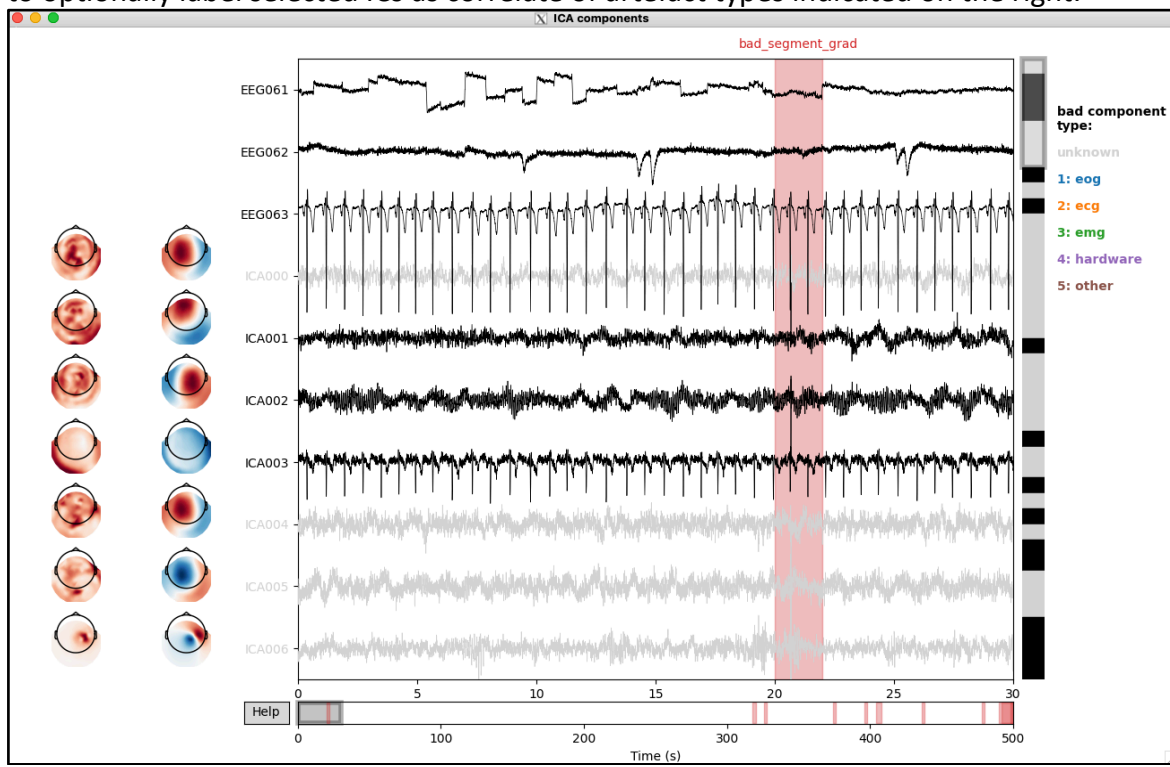

Figure 4. Interactive labelling of independent components (ICs) using the `osl_ica_label` tool. The weights and time courses for each IC are shown as rows. Bad ICs are indicated as coloured the time courses (i.e., other than black, see types on the right); annotations in the time courses indicate bad segments.

Further inspecting the summary and subject report also reveals that the automatic bad IC detection did not identify any EOG-related components and identified spurious ECG-related components in subject 19, and similarly in session 5 of subject 10. The ICA labels of these sessions are also adapted using the same interactive GUI. Once the user is satisfied with the rejected components, the following command line function is used to iteratively remove all the selected components from the data, which will also automatically update the logs and reports:

```
(osle) > osl_ica_apply processed
```

### 3.4 Coregistration and source reconstruction

Listing 3 shows an example script that carries out coregistration and source reconstruction. This extracts the head shape information from the `.fif` file, and specifies a custom function that removes all head shape points below the fiducials (lower head shape points extend beyond the sMRI file, i.e., in the neck area, which will compromise the coregistration). This also computes the surfaces of the inside and outside of the skull, and the scalp using FSL tools (e.g. FLIRT and BET); and coregisters the sMRI, headshape and MEG data, and computes the forward model using a single shell Boundary Element Model (BEM). Volumetric LCMV beamforming (Van Veen et al., 1997) is used to estimate source activity on an 8 mm volumetric source grid, and combine source dipoles into a 52 parcels by estimating a spatial basis set over all dipole locations within each parcel, and reduces spatial leakage between the parcels (Colclough et al., 2015).

Figure 5 shows the output directory structure after running the script in Listing 3. Most importantly, the `subxxx-runxx_parc-raw.fif` file contains the parcellated and orthogonalized data in MNE-Python `Raw` format. Because the sign of source reconstructed data are ambiguous, we align the signs for each parcel over subjects, using one subject as a template, see Listing 4. This additionally generates the `subxxx-runxx_sflip_parc-raw.fif` file, which we will use for statistical analysis below. `run_src_batch` also generated the source *subject* and *summary* HTML reports (Figure 6), which can be used for quality assurance.

```python
import os
import numpy as np
from dask.distributed import Client
from osl_ephys import source_recon, utils

source_recon.setup_fsl("~/fsl") # FSL needs to be installed

def fix_headshape_points(outdir, subject):
    filenames = source_recon.rhino.get_coreg_filenames(outdir, subject)

    # Load saved headshape and nasion files
    hs = np.loadtxt(filenames["polhemus_headshape_file"])
    nas = np.loadtxt(filenames["polhemus_nasion_file"])
    lpa = np.loadtxt(filenames["polhemus_lpa_file"])
    rpa = np.loadtxt(filenames["polhemus_rpa_file"])

    # Remove headshape points on the nose
    remove = np.logical_and(hs[1] > max(lpa[1], rpa[1]), hs[2] < nas[2])
    hs = hs[:, ~remove]

    # Overwrite headshape file
    utils.logger.log_or_print(f"overwritting {filenames['polhemus_headshape_file']}")
    np.savetxt(filenames["polhemus_headshape_file"], hs)

if __name__ == "__main__":
    utils.logger.set_up(level="INFO")
    client = Client(n_workers=16, threads_per_worker=1)

    config = """
        source_recon:
        - extract_polhemus_from_info: {}
        - fix_headshape_points: {}
        - compute_surfaces:
            include_nose: False
        - coregister:
            use_nose: False
            use_headshape: True
        - forward_model:
            model: Single Layer
        - beamform_and_parcellate:
            freq_range: [1, 45]
            chantypes: [mag, grad]
            rank: {meg: 60}
            parcellation_file: Glasser52_binary_space-MNI152NLin6_res-8x8x8.nii.gz
            method: spatial_basis
            orthogonalisation: symmetric
    """

    basedir = "ds117"
    proc_dir = os.path.join(basedir, "processed")

    # Define inputs
    subjects = [f"sub{i+1:03d}-run{j+1:02d}" for i in range(19) for j in range(6)]
    preproc_files = sorted(utils.Study(os.path.join(proc_dir, "sub{sub_id}-
        run{run_id}/sub{sub_id}-run{run_id}_preproc-raw.fif")).get())
    smri_files = np.concatenate([[smri_file]*6 for smri_file in
        sorted(utils.Study(os.path.join(basedir,
        "sub{sub_id}/anatomy/highres001.nii.gz"))).get()])

    # Run source batch
    source_recon.run_src_batch(
        config,
        outdir=proc_dir,
        subjects=subjects,
        preproc_files=preproc_files,
        smri_files=smri_files,
        extra_funcs=[fix_headshape_points],
        dask_client=True,
    )
```

Listing 3. Example coregistration and source reconstruction script.

```python
import os
from glob import glob
from dask.distributed import Client

from osl_ephys import source_recon, utils

source_recon.setup_fsl("~/fsl")

# Directory containing source reconstructed data
proc_dir = "ds117/processed"
src_files = sorted(utils.Study(os.path.join(proc_dir,
    "sub{sub_id}-run{run_id}/parc/parc-raw.fif")).get())

if __name__ == "__main__":
    utils.logger.set_up(level="INFO")

    subjects = [f"sub{i+1:03d}-run{j+1:02d}" for i in range(19) for j in range(6)]

    # Find a good template subject to match others to
    template = source_recon.find_template_subject(
        proc_dir, subjects, n_embeddings=15, standardize=True,
    )

    # Settings
    config = f"""
        source_recon:
        - fix_sign_ambiguity:
            template: {template}
            n_embeddings: 15
            standardize: True
            n_init: 3
            n_iter: 3000
            max_flips: 20
    """

    # Setup parallel processing
    client = Client(n_workers=16, threads_per_worker=1)

    # Run sign flipping
    source_recon.run_src_batch(config, proc_dir, subjects, dask_client=True)
```

Listing 4. Example sign-flipping script.

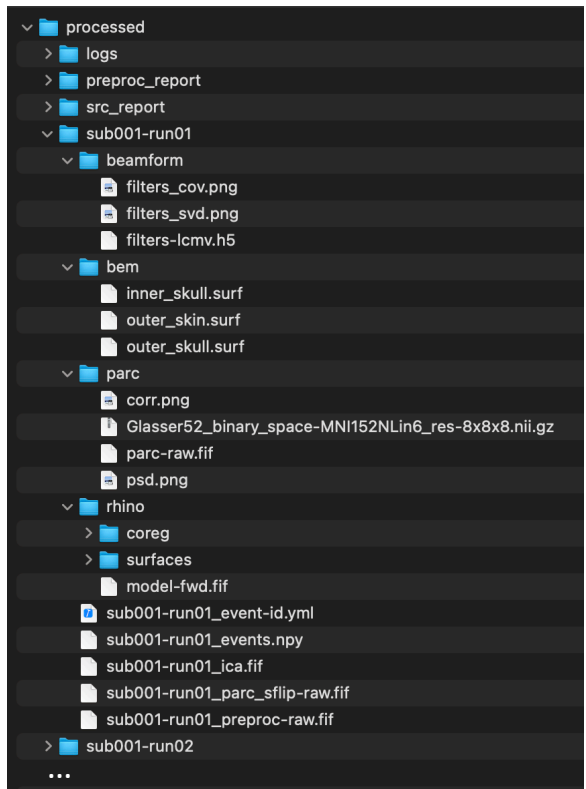

Figure 5. The output directory structure of `run_src_batch`. All outputs are saved in the same general directory and subject/session directories that contain the previously preprocessed data.

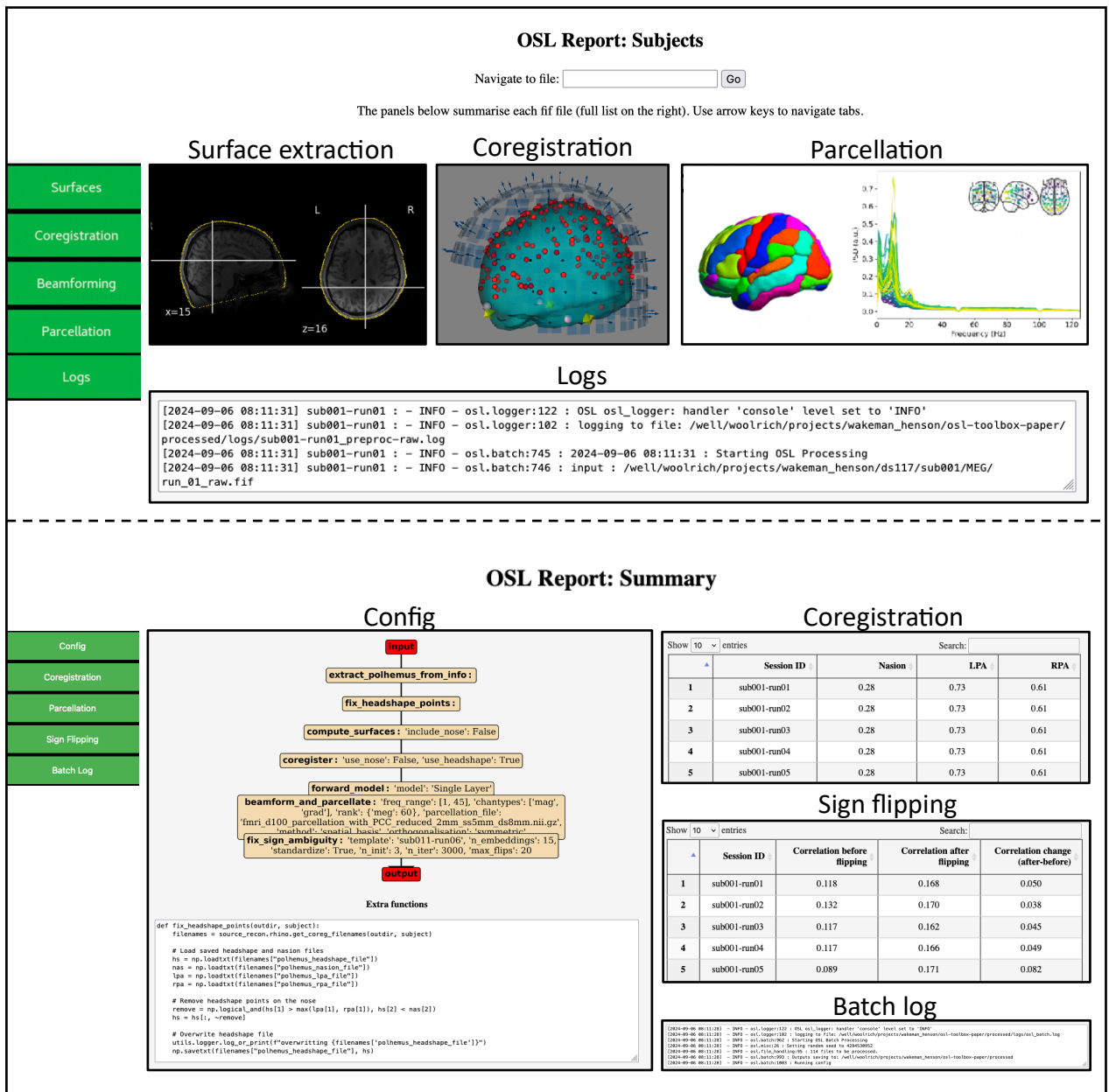

Figure 6. Example of the `source_recon` subject (top) and summary (bottom) report. The subject report contains figures showing the extracted surfaces, the parcel power spectra, and interactive figures showing the coregistration. The summary report contains summary information and quantitative metrics of all files processed using `run_src_batch`, including interactive tables that can guide the user to specific subjects/sessions which might require further attention.

### 3.5 Epoching and statistical analysis using general linear models (GLM)

Listing 5 shows an example script to epoch the source reconstructed data and perform statistical analysis. The data will be epoched around the onset of the visual stimulus using an MNE-Python method and fit a first level GLM using a custom function, `first_level`, which sets up three regressors for the three different stimulus types: *famous*, *unfamiliar*, and *scrambled* faces, and adds a *mean* and *faces vs. scrambled* contrast.

The `preproc` section of the config is used to specify the use of epoching and the custom `first_level` function (see Listing 5). The `read_dataset` function with option `ftype: sflip_parc-raw` will load in the previously generated data objects, where the sign-flipped, parcellated data are loaded into `dataset[`raw`]`. It is necessary to specify `overwrite=True` in `run_proc_batch` to ensure processing takes place (i.e., the default option is `overwrite=False`, which will skip processing if the data represented in `dataset['raw']` already exsists on disk). However, these data are only loaded to provide to downstream functions and thus do not need to actually be overwritten. Therefore, the `skip_save` option is used to skip saving the data just loaded in with `read_dataset`. Now, only the newly generated data in the dataset (`epochs` and `glm`) will be saved (user defined items in the `dataset` are saved as `.pkl` files).

The batch processing function will first loop over subjects/sessions to run the functions in the `preproc` field of the `config`. After all sessions are processed, it will run the functions in the `group` field of the `config` on all the `preproc` outputs. A design matrix with a regressor for each subject is constructed along with a group mean contrast. Finally, maximum statistic permutation test is used to test whether there is a difference between normal face stimuli, and scrambled faces in the 50-300 ms post stimulus onset window. The script in Listing 5 generates the output figures in Figure 7.

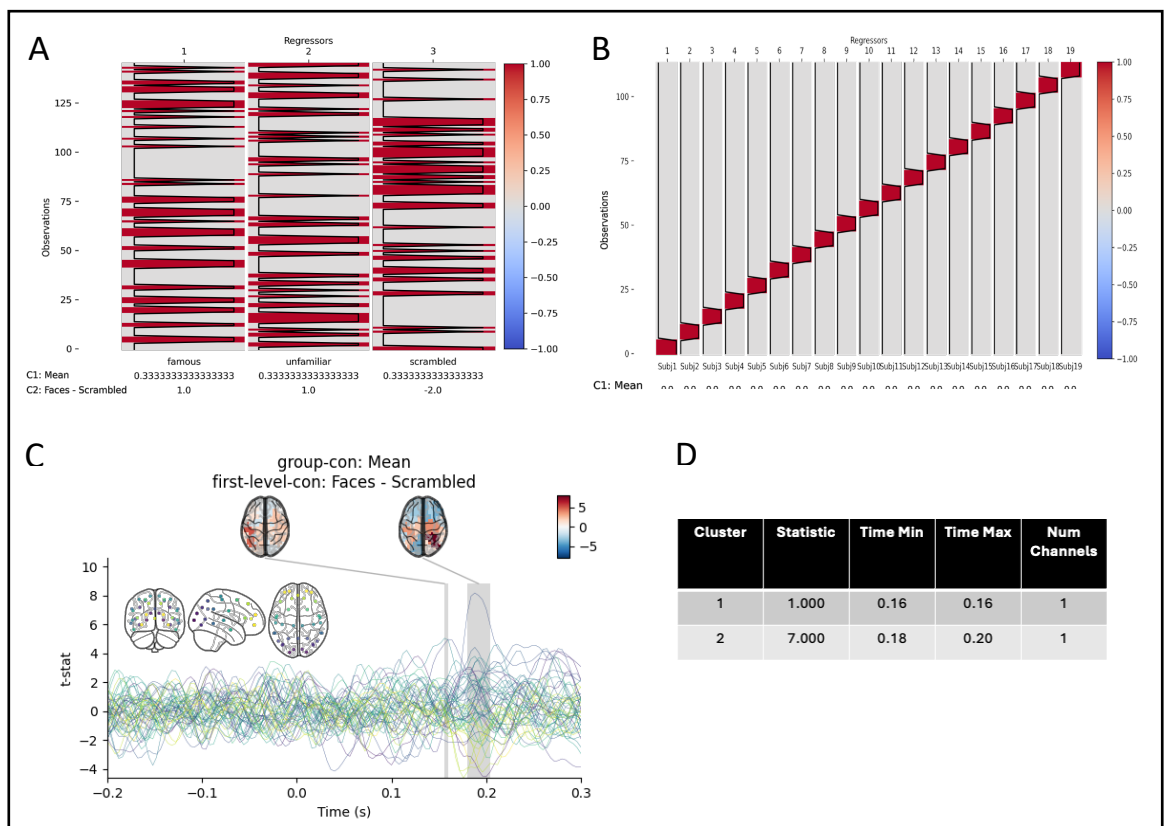

Figure 7. Pictures of real faces elicit statistically different event-related fields than pictures of scrambled faces. A) example of a first level (session) design matrix, with three regressors and two contrasts. B) The second level (group) design contains 19 subject regressors, and one mean contrast. C) The group *Faces – Scrambled* contrast. Coloured lines show individual parcels, with colours in an anterior-posterior gradient (inset). Shaded areas show significant

time periods, and topographies the mean t-statistic in each period. D) Extent of significant times periods.

```python
import os
import numpy as np
import glmtools
import matplotlib.pyplot as plt
from dask.distributed import Client
from osl_ephys import preprocessing, glm

def first_level(dataset, userargs):
    DC = glmtools.design.DesignConfig()
    DC.add_regressor(name="famous", rtype="Categorical", codes=[5,6,7])
    DC.add_regressor(name="unfamiliar", rtype="Categorical", codes=[13,14,15])
    DC.add_regressor(name="scrambled", rtype="Categorical", codes=[17,18,19])
    DC.add_contrast(name="Mean", values={"famous": 1/3, "unfamiliar": 1/3, "scrambled": 1/3})
    DC.add_contrast(name="Faces - Scrambled", values={"famous": 1, "unfamiliar": 1,
        "scrambled": -2})
    dataset['glm'] = glm.glm_epochs(DC, dataset['epochs'])
    dataset['glm'].design.plot_summary(savepath=os.path.join(
        os.path.dirname(dataset['raw'].filenames[0]), 'subject_design.png'))
    return dataset

def second_level(dataset, userargs):
    firstlevel_contrast = userargs.get('firstlevel_contrast', 'Faces - Scrambled')
    group_contrast = userargs.get('group_contrast', 'Mean')
    tmin = userargs.get('tmin', -np.Inf)
    tmax = userargs.get('tmax', np.Inf)

    groupDC = glmtools.design.DesignConfig()
    info = {"Subject": np.repeat(np.arange(1, 20), 6)}
    for i in range(19):
        # Add subject mean regressors
        groupDC.add_regressor(name=f"Subj{i+1}", rtype="Categorical", datainfo="Subject",
            codes=[i+1])

    # add group contrast
    groupDC.add_contrast(name='Mean', values={f"Subj{i+1}": 1/19 for i in range(19)})

    # group level model
    dataset['group_glm'] = glm.group_glm_epochs(dataset['glm'], groupDC)
    dataset['group_glm'].design.plot_summary(savepath=os.path.join(userargs['figdir'],
        'group_design.png'))

    # max stat permutation test
    dataset['group_glm_perm'] = glm.glm_base.SensorMaxStatPerm(dataset['group_glm'],
        dataset['group_glm'].contrast_names.index(group_contrast),
        dataset['group_glm'].fl_contrast_names.index(firstlevel_contrast), tmin=tmin, tmax=tmax)
    dataset['group_glm_perm'].plot_sig_clusters(99)
    plt.savefig(os.path.join(userargs['figdir'], f'group_contrast-{firstlevel_contrast}-
        significance.png'))
    return dataset

if __name__ == "__main__":
    client = Client(n_workers=16, threads_per_worker=1)

    config = """
      preproc:
        - read_dataset: {ftype: sflip_parc-raw}
        - epochs: {picks: misc, tmin: -0.2, tmax: 0.5}
        - first_level: {}
      group:
        - second_level: {tmin: 0.05, tmax: 0.3, figdir: ds117/figures}
    """

    proc_dir = "ds117/processed"
    src_files = sorted(utils.Study(os.path.join(proc_dir,
        "sub{sub_id}-run{run_id}", "sub{sub_id}-run{run_id}_sflip_parc-raw.fif")).get())
    subjects = [f"sub{i+1:03d}-run{j+1:02d}" for i in range(19) for j in range(6)]

    preprocessing.run_proc_batch(
        config,
        src_files,
        subjects,
        outdir=proc_dir,
        ftype='raw',
        extra_funcs=[first_level, second_level],
        dask_client=True,
        overwrite=True,
        gen_report=False,
        skip_save=['events', 'raw', 'ica', 'event_id', 'sflip_parc-raw'],
    )
```

Listing 5. An example script for epoching, first and second level GLM, and permutation testing.

## 4. Discussion

We have presented how the `osl-ephys` toolbox for the analysis of M/EEG data. This is not a standalone toolbox, but heavily relies on the widely adopted MNE-Python toolbox, FSL and other popular python packages: numpy, scipy, matplotlib, etc. `osl-ephys` aims to augment MNE-Python by providing a config API for reproducible processing of large quantities of data, while providing quality assurance and unique functionalities for data analysis. This include functions for automatic and manual data preprocessing, FSL-based (Freesurfer independent) volumetric source reconstruction, and statistical analysis, in particular, using GLMs.

Researchers face a number of challenges when analysing M/EEG data. Firstly, analysis is complex and heterogenous. The analysis pipeline depends on the nature and quality of the data, as well as the experimental design and research question. Therefore, analysis flexibility is essential for analysis software. However, analysis complexity and flexibility come with caveats, particularly in terms of transparency and reproducibility. In particular, it is cumbersome and error-prone to manually provide *all* details of an analysis pipeline in the Methods section of academic publications. Even with the growing requirement of funders and journals to provide analysis scripts upon publication of a manuscript, the full details for the analysis pipeline often remain unclear.

Therefore, `osl-ephys` uses a concise and easily shareable `config` API, which reduces the amount of custom written scripts and functions that the researchers need to write (whilst retaining analysis flexibility). In addition, `osl-ephys` keeps track of all processing that took place in log files, and it generates analysis reports that can be used for both reproducibility efforts, and quality assurance.

The high complexity also means that no single analysis toolbox can provide all possible analytical methods, and therefore, the researcher typically needs to stitch together various third-party toolboxes in their analysis pipeline. `osl-ephys` is built on top of the most adopted Python M/EEG analysis toolbox (MNE-Python), and many other Python (and MATLAB) toolboxes contain plugins and/or documentation on how to use their toolbox in combination with MNE-Python. This makes it more straightforward to use different toolboxes. Additionally, `osl-ephys` can be used in a modular fashion, and custom-written and third-party functionality can be easily implemented as an extension to `osl-ephys` by supplying the *chain*/*batch* functions with extra function definitions.

Another challenge is that high analysis complexity means a high entrance barrier for new researchers in the field of M/EEG analysis, and/or programming, especially considering the multidisciplinary nature of the field. `osl-ephys` alleviates this by combining the `config` API a limited number of functions (in particular the *chain* and *batch* functions) that the user interacts with and taking care of much of the complexity in programming and data bookkeeping on the backend. This is further aided by comprehensive documentation and tutorials. Finally, the analysis reports can also help researchers new to the field, by providing a platform for quality assurance.

## 5. Acknowledgements

This research was supported by the National Institute for Health Research (NIHR) Oxford Health Biomedical Research Centre. The Wellcome Centre for Integrative Neuroimaging is supported by core funding from the Wellcome Trust (203139/Z/16/Z). M.V.E. is supported by the Wellcome Trust (106183/Z/14/Z, 215573/Z/19/Z), the New Therapeutics in Alzheimer's Diseases (NTAD) and Synaptic Health in Neurodegeneration (SHINE) studies supported by the MRC and the Dementia Platform UK (RG94383/RG89702). C.G. is supported by the Wellcome Trust (215573/Z/19/Z). M.W.W. is supported by the Wellcome Trust (106183/Z/14/Z, 215573/Z/19/Z), the New Therapeutics in Alzheimer's Diseases (NTAD) and Synaptic Health in Neurodegeneration (SHINE) studies supported by UK MRC, the Dementia Platform UK (RG94383/RG89702), and the NIHR Oxford Health Biomedical Research Centre (NIHR203316). The views expressed are those of the author(s) and not necessarily those of the NIHR or the Department of Health and Social Care.

## 6. Author Contributions

M.V.E.: Conceptualisation; Methodology; Software; Validation; Formal Analysis; Investigation; Resources; Data Curation; Writing—Original Draft; Writing—Reviewing & Editing; Visualisation; Project administration. C.G.: Conceptualisation; Methodology; Software; Validation; Formal Analysis; Investigation; Resources; Data Curation; Writing—Reviewing & Editing. A.J.Q.: Conceptualisation; Methodology; Software; Validation; Formal Analysis; Investigation; Resources; Data Curation; Writing—Reviewing & Editing. M.W.W.: Conceptualisation; Methodology; Software; Validation; Writing—Reviewing & Editing; Supervision; Funding Acquisition.

## 7. References

Ågren, W. (2023). Neurocode. Available at: https://github.com/neurocode-ai/neurocode/tree/main?tab=MIT-1-ov-file

Brodbeck, C., Das, P., Brooks, T. L., Reddigari, S., and jpkulasingham (2023). Eelbrain. doi: 10.5281/ZENODO.7951251

Colclough, G. L., Brookes, M. J., Smith, S. M., and Woolrich, M. W. (2015). A symmetric multivariate leakage correction for MEG connectomes. *NeuroImage* 117, 439–448. doi: 10.1016/j.neuroimage.2015.03.071

Dask Development Team (2016). Dask: Library for dynamic task scheduling. Available at: http://dask.pydata.org

Delorme, A., and Makeig, S. (2004). EEGLAB: an open source toolbox for analysis of single-trial EEG dynamics including independent component analysis. *Journal of Neuroscience Methods* 134, 9–21. doi: 10.1016/j.jneumeth.2003.10.009

Gohil, C., Huang, R., Roberts, E., Van Es, M. W. J., Quinn, A. J., Vidaurre, D., et al. (2023). osl-dynamics: A toolbox for modelling fast dynamic brain activity. Neuroscience. doi: 10.1101/2023.08.07.549346

Gramfort, A. (2013). MEG and EEG data analysis with MNE-Python. *Front. Neurosci.* 7. doi: 10.3389/fnins.2013.00267

Henson, R. (2024). Guideline(s) for MEG Pre-processing. Available at: https://imaging.mrc-cbu.cam.ac.uk/meg/maxpreproc (Accessed July 8, 2024).

Jas, M., Thorpe, R., Tolley, N., Bailey, C., Brandt, S., Caldwell, B., et al. (2023). HNN-core: A Python software for cellular and circuit-level interpretation of human MEG/EEG. doi: 10.5281/ZENODO.10288598

Jenkinson, M., Beckmann, C. F., Behrens, T. E. J., Woolrich, M. W., and Smith, S. M. (2012). FSL. *NeuroImage* 62, 782–790. doi: 10.1016/j.neuroimage.2011.09.015

Kohl, O., Woolrich, M., Nobre, A. C., and Quinn, A. (2023). Glasser52: A parcellation for MEG-Analysis. doi: 10.5281/ZENODO.10401793

Larson, E., Gramfort, A., Engemann, D. A., Leppakangas, J., Brodbeck, C., Jas, M., et al. (2023). MNE-Python. doi: 10.5281/ZENODO.592483

Litvak, V., Mattout, J., Kiebel, S., Phillips, C., Henson, R., Kilner, J., et al. (2011). EEG and MEG Data Analysis in SPM8. *Computational Intelligence and Neuroscience* 2011, 1–32. doi: 10.1155/2011/852961

Lopez-Calderon, J., and Luck, S. J. (2014). ERPLAB: an open-source toolbox for the analysis of event-related potentials. *Front. Hum. Neurosci.* 8. doi: 10.3389/fnhum.2014.00213

Lu, Z. (2020). PyCTRSA: A Python package for cross-temporal representational similarity analysis-based E/MEG decoding. doi: 10.5281/ZENODO.4273673

OHBA Analysis Group (2014). OSL MATLAB. Available at: https://github.com/OHBA-analysis/osl

Oostenveld, R., Fries, P., Maris, E., and Schoffelen, J.-M. (2011). FieldTrip: Open Source Software for Advanced Analysis of MEG, EEG, and Invasive Electrophysiological Data. *Computational Intelligence and Neuroscience* 2011, 1–9. doi: 10.1155/2011/156869

Palva, J. M., Wang, S. H., Palva, S., Zhigalov, A., Monto, S., Brookes, M. J., et al. (2018). Ghost interactions in MEG/EEG source space: A note of caution on inter-areal coupling measures. *NeuroImage* 173, 632–643. doi: 10.1016/j.neuroimage.2018.02.032

Poldrack, R. A., and Gorgolewski, K. J. (2017). OpenfMRI: Open sharing of task fMRI data. *NeuroImage* 144, 259–261. doi: 10.1016/j.neuroimage.2015.05.073

Quinn, A., and Hymers, M. (2020). SAILS: Spectral Analysis In Linear Systems. *JOSS* 5, 1982. doi: 10.21105/joss.01982

Quinn, A. J. (2019). glmtools. Available at: https://gitlab.com/ajquinn/glmtools

Quinn, A. J., Atkinson, L. Z., Gohil, C., Kohl, O., Pitt, J., Zich, C., et al. (2024). The GLM-spectrum: A multilevel framework for spectrum analysis with covariate and confound modelling. *Imaging Neuroscience* 2, 1–26. doi: 10.1162/imag_a_00082

Rosner, B. (1983). Percentage Points for a Generalized ESD Many-Outlier Procedure. *Technometrics* 25.

Sabbagh, D., Ablin, P., Varoquaux, G., Gramfort, A., and Engemann, D. A. (2020). Predictive regression modeling with MEG/EEG: from source power to signals and cognitive states. *NeuroImage* 222, 116893. doi: 10.1016/j.neuroimage.2020.116893

Schirrmeister, R. T., Springenberg, J. T., Fiederer, L. D. J., Glasstetter, M., Eggensperger, K., Tangermann, M., et al. (2017). Deep learning with convolutional neural networks for EEG decoding and visualization. *Human Brain Mapping* 38, 5391–5420. doi: 10.1002/hbm.23730

Tadel, F., Baillet, S., Mosher, J. C., Pantazis, D., and Leahy, R. M. (2011). Brainstorm: A User-Friendly Application for MEG/EEG Analysis. *Computational Intelligence and Neuroscience* 2011, 1–13. doi: 10.1155/2011/879716

Taulu, S., Simola, J., and Kajola, M. (2005). Applications of the signal space separation method. *IEEE Trans. Signal Process.* 53, 3359–3372. doi: 10.1109/TSP.2005.853302

The MathWorks Inc. (2020). MATLAB.

Tzourio-Mazoyer, N., Landeau, B., Papathanassiou, D., Crivello, F., Etard, O., Delcroix, N., et al. (2002). Automated Anatomical Labeling of Activations in SPM Using a Macroscopic Anatomical Parcellation of the MNI MRI Single-Subject Brain. *NeuroImage* 15, 273–289. doi: 10.1006/nimg.2001.0978

Van Veen, B. D., van Drongelen, W., Yuchtman, M., and Suzuki, A. (1997). Localization of brain electrical activity via linearly constrained minimum variance spatial filtering. *IEEE Transactions on Biomedical Engineering* 44, 867–880. doi: 10.1109/10.623056

Wakeman, D. G., and Henson, R. N. (2015). A multi-subject, multi-modal human neuroimaging dataset. *Sci Data* 2, 150001. doi: 10.1038/sdata.2015.1